\documentclass[letter]{aa} % for the letters
\usepackage{graphicx}
\usepackage{txfonts}
\usepackage{natbib}
\bibpunct{(}{)}{;}{a}{}{,}

\usepackage{lscape}
\usepackage{longtable}
\usepackage{amssymb}
\usepackage{xfrac}
\usepackage{color}

\begin{document}

% Version 1
   \title{Protostellar Outflows at the EarliesT Stages (POETS)}
%  \subtitle{II. Diverse evolutionary stages in HMSFR traced by radio continuum}
   \subtitle{II. A possible radio synchrotron jet associated with the EGO G035.02$+$0.35}

%  \author{A. Sanna\inst{1}  et al.}
  
   \author{A.~Sanna\inst{1,2} \and L.~Moscadelli\inst{3} \and C.~Goddi\inst{4} \and M.~Beltr\'{a}n\inst{3} \and C.\,L.~Brogan\inst{5} 
               A.~Caratti~o~Garatti\inst{6}  \and C.~Carrasco-Gonz\'{a}lez\inst{7} \and T.\,R.~Hunter\inst{5} \and F.~Massi\inst{3} 
               \and M.~Padovani\inst{3}}
            
%\fnmsep\thanks{Just to show the usage of the elements in the author field}

   \institute{Max-Planck-Institut f\"{u}r Radioastronomie, Auf dem H\"{u}gel 69, 53121 Bonn, Germany \\
   \email{asanna@mpifr-bonn.mpg.de}
   \and INAF, Osservatorio Astronomico di Cagliari, via della Scienza 5, 09047 Selargius (CA), Italy   
   \and INAF, Osservatorio Astrofisico di Arcetri, Largo E. Fermi 5, 50125 Firenze, Italy
   \and Department of Astrophysics/IMAPP, Radboud University Nijmegen, PO Box 9010, NL-6500 GL Nijmegen, the Netherlands
   \and NRAO, 520 Edgemont Road, Charlottesville, VA 22903, USA
   \and Dublin Institute for Advanced Studies, Astronomy \& Astrophysics Section, 31 Fitzwilliam Place, Dublin 2, Ireland
   \and Instituto de Radioastronom\'{i}a y Astrof\'{i}sica UNAM, Apartado Postal 3-72 (Xangari), 58089 Morelia, Michoac\'{a}n, M\'{e}xico}    
   
   \date{Received 1 November 2018; accepted 23 January 2019}

% \abstract{}{}{}{}{}
% 5 {} token are mandatory

  \abstract{Centimeter continuum observations of protostellar jets have revealed the presence of knots of shocked gas where the flux density
  decreases with frequency. This spectrum is characteristic of nonthermal synchrotron radiation and implies the presence of both magnetic fields
  and relativistic electrons in protostellar jets. Here, we report on one of the few detections of nonthermal jet driven by a young massive
  star in the star-forming region G035.02$+$0.35. We made use of the NSF's Karl G. Jansky Very Large Array (VLA) to observe this region at C, Ku, and
  K bands with the A- and B-array configurations, and obtained sensitive radio continuum maps down to a rms of 10\,$\mu$Jy\,beam$^{-1}$. These
  observations allow for a detailed spectral index analysis of the radio continuum emission in the region, which we interpret as a protostellar jet with
  a number of knots aligned with extended 4.5\,$\mu$m emission. Two knots clearly emit nonthermal radiation and are found at similar distances, of approximately
  10,000\,au, each side of the central young star, from which they expand at velocities of hundreds km\,s$^{-1}$. We estimate both the
  mechanical force and the magnetic field associated with the radio jet, and infer a lower limit of $0.4\times10^{-4}$\,M$_{\odot}$\,yr$^{-1}$\,km\,s$^{-1}$
  and values in the range 0.7--1.3\,mG, respectively.} 
  
  % context heading (optional),  leave it empty if necessary


  \titlerunning{The nonthermal radio jet  associated with the EGO G035.02$+$0.35}
   \maketitle
%
%________________________________________________________________

\section{Introduction}

The centimeter continuum emission of ionized protostellar jets is a preferred tracer to image the outflow geometry, and, knowing the ionization fraction of
local hydrogen gas, to quantify the mass-loss rate at scales from a few 10s to 1000s\,au of the central young star \citep[e.g.,][]{Reynolds1986,Tanaka2016}.
Radio jet properties have been recently reviewed by \citet{Anglada2018}. The flux density ($S_{\nu}$) of radio jets typically increases with frequency ($\nu$) 
within a few 1000s\,au of the central star ($S_{\nu}\propto\nu^{\alpha}$), showing a partially opaque spectral index ($0<\alpha<2$).
These spectra are interpreted as thermal free-free emission from ionized particles accelerated within their own electric field. Notably, along the axis of a few
radio jets, the spectrum is occasionally inverted at the loci of bright knots of ionized gas \citep[e.g.,][]{Rodriguez2005,Carrasco2010,Moscadelli2013,RodriguezKamenetzky2016,Osorio2017}, showing spectral index values much less than the optically thin limit ($-0.1$). 
These spectra are interpreted as evidence for (nonthermal) synchrotron emission from strong jet shocks against the ambient medium, where ionized particles
would be sped up to relativistic velocities via diffusive shock acceleration \citep[e.g.,][]{Padovani2015,Padovani2016}. 

For the prototypical synchrotron jet HH\,80--81, \citet{Carrasco2010} detected  linear polarization of the centimeter continuum emission for the first time, proving 
that the negative slope of the radio spectrum is due to synchrotron radiation. Polarized maser emission associated with protostellar outflows provides further evidence
for the presence of magnetic fields, and observations of maser cloudlets at milli-arcsecond resolution show a strong correlation between the local orientation of proper
motion and magnetic field vectors \citep[e.g.,][]{Surcis2013,Sanna2015,Goddi2017,Hunter2018}. In this Letter, we report on one of the very few detections of a
nonthermal jet emitted by a young massive star, discovered in the star-forming region G035.02$+$0.35 as part of the Protostellar Outflow at the EarliesT Stage
(POETS) survey \citep[][hereafter Paper\,I]{Moscadelli2016,Sanna2018}.

%_____________________________________________________________
%                                              FIGURES  N.1
%-----------------------------------------------------------------------------------------------------------
\begin{figure*}
%\sidecaption
%\centering
\includegraphics [angle= 0, scale= 0.9]{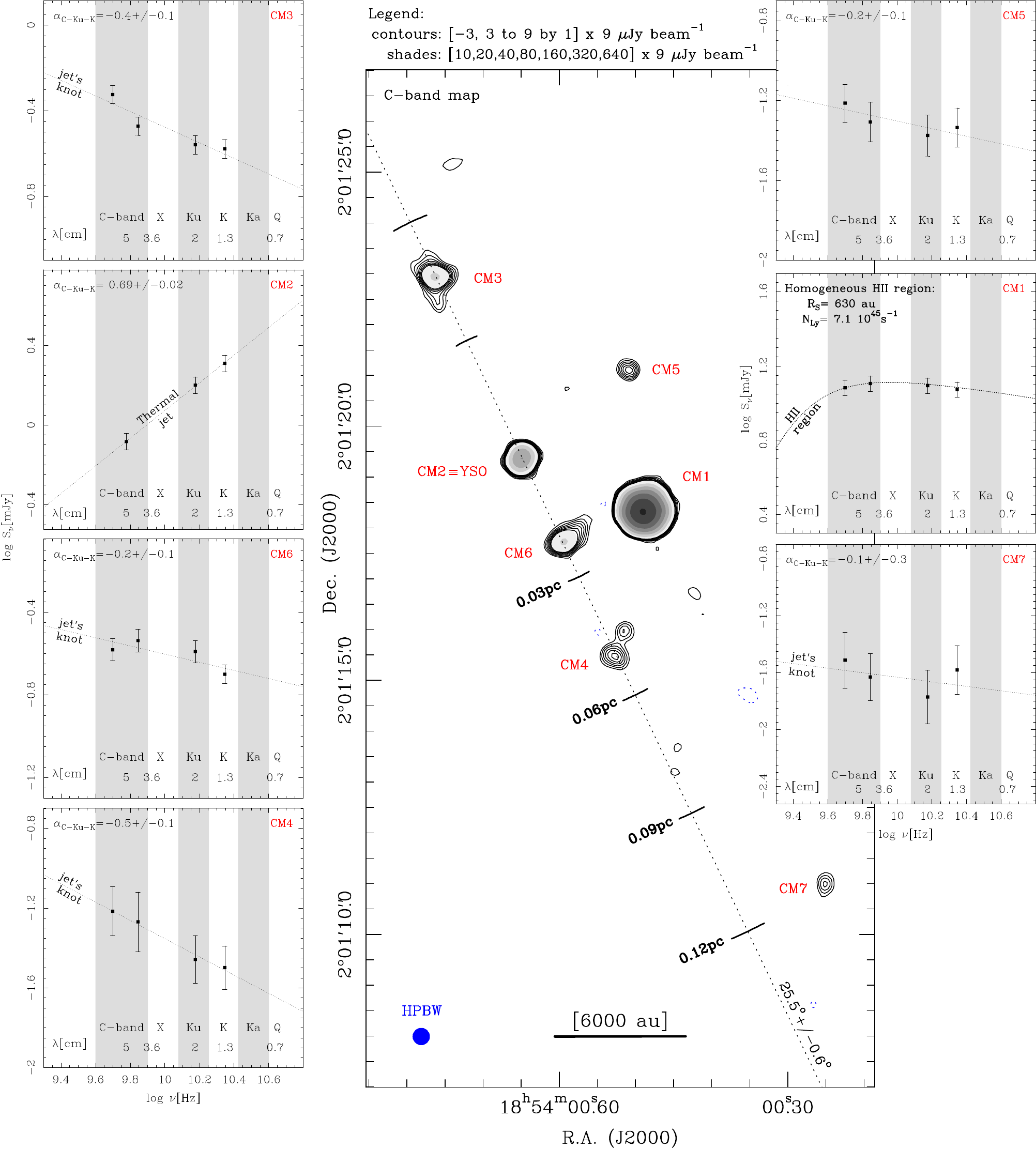}
\caption{Extended radio jet emission driven by source CM2 in G035.02$+$0.35.
\textbf{Middle Panel:} continuum map obtained with the A-configuration of the VLA at C-band (without constraints on \textit{uv}-coverage). Contour levels, at multiple of 
the 1\,$\sigma$ rms, are indicated on top; levels higher than 10\,$\sigma$ are drawn in grey shades. The beam size is shown in
the bottom left corner. A scale bar in units of au is drawn near the bottom axis. The field-of-view covers the radio continuum sources
detected within 0.1\,pc from CM2 (CM8 in Fig.\ref{fig3}), whose centimeter spectral index ($\alpha$) traces radio thermal jet emission
(from Paper\,I). Labels are indicated in red near each source. Components CM3, CM4, and CM6 are aligned along a position angle of 
25.5$\degr$\,$\pm$\,0.6$\degr$ with respect to CM2 (dotted line), tracing the location of four jet's knots. Ticks are marked along
the jet direction at steps of 6000\,au from CM2.   
\textbf{Side Panels:} centimeter spectral energy distribution of radio components surrounding CM2. Each plot shows the logarithm
of the integrated flux (in mJy) as a function of the logarithm of the observed frequency (in Hz) for four data points. Details for CM2 were presented 
in  Paper\,I. Different frequency bands are labeled near the bottom axis together with the reference wavelength;
grey shades mark the boundary of each band.
The spectral index values ($\alpha$) and their uncertainties, derived from a linear fit of the four data points, are specified in the upper left corner of each
panel. Assuming an electron temperature of 10$^4$\,K, the spectral energy distribution of source CM1 is well reproduced by a model of
homogeneous H\,{\sc ii} region with: Str\"{o}mgren radius (R$_S$) of 630\,au, number of Lyman photons (N$_{Ly}$) of 10$^{45.85}$\,s$^{-1}$,
electron density (n$_e$) of 8.8\,$\times$\,10$^{4}$\,cm$^{-3}$, and emission measure (EM) of 4.8\,$\times$\,10$^{7}$\,pc\,cm$^{-6}$.}\label{fig1}
\end{figure*}

%_____________________________________________________________
%-----------------------------------------------------------------------------------------------------------

The star-forming region G035.02$+$0.35 hosts diverse stages of stellar evolution, including a hot molecular core (HMC) near to a hyper compact (HC)
H\,{\sc ii} region (e.g., \citealt{Brogan2011,Beltran2014}). At a parallax distance of 2.33\,kpc from the Sun \citep{Wu2014}, the entire region emits
a bolometric luminosity of 1--3\,$\times$\,10$^4$\,L$_{\odot}$ (\citealt{Beltran2014}; Towner et al.\,2018, in prep.). G035.02$+$0.35 was surveyed
with the IRAC instrument onboard of Spitzer and classified as an extended green object (EGO) associated with bright 4.5\,$\mu$m emission, which is a
tracer of ambient gas shocked by early outflow activity \citep{Cyganowski2008,Cyganowski2009,Lee2013}. \citet{Cyganowski2011} made a census of
the 8\,GHz continuum sources at the center of the IR nebula with the B-configuration of the Very Large Array (VLA). They identified
5 compact radio sources above a threshold of 100\,$\mu$Jy\,beam$^{-1}$. Here, we make use of the longest VLA baselines at 6, 15, and 22\,GHz (C,
Ku, and K bands, respectively) to image the continuum emission at a sensitivity of 10\,$\mu$Jy\,beam$^{-1}$. Observation information was presented
in Paper\,I and is summarized in Table\,\ref{settings}. We have performed a detailed spectral index analysis for each radio continuum component and we have
identified an extended radio jet with nonthermal knots. This radio jet is driven by the HMC source at the base of the IR nebula.

%_____________________________________________________________
%                                                    TABLES  # 1
%-----------------------------------------------------------------------------------------------------------

\onltab{
\begin{table*}
\caption{Summary of VLA observations towards G035.02$+$0.35 (code 14A-133).}\label{settings}
\centering
\begin{tabular}{c c c c c c}

\hline \hline

 Band   &  $\nu_{\rm center}$   &  BW     &  Array  &  HPBW  &            RMS noise                   \\  
           &             (GHz)             &  (GHz)  &           &   ($''$)  &  ($\mu$Jy\,beam$^{-1}$)     \\  

\hline
 C   & 6.0   & 4.0 & A  &  0.338 &  9.0  \\
 Ku & 15.0 & 6.0 & A  &  0.138 &  9.0  \\
 K   & 22.2 & 8.0 & B  &  0.308 &  9.0  \\

%& & & & & & & & &  \\
\hline
%& & & & & & & & &  \\
\end{tabular}

\tablefoot{Columns\,1, 2, and 3: radio band, central frequency of the observations, and receiver bandwidth used. Columns\,4 and 5: array configuration
and synthesized beam size of the observations. Column\,6: actual rms noise of the maps. Observations were conducted on 2014 March 12 (Ku band),
May 12 (C band), and on 2015 February 14 (K band). Further details can be found in Paper\,I.}
\end{table*}
%\end{landscape}
}
%_____________________________________________________________
%-----------------------------------------------------------------------------------------------------------

%_____________________________________________________________
%                                                    TABLES # 2
%-----------------------------------------------------------------------------------------------------------
%\addtocounter{table}{0}
%\begin{landscape}
\onltab{
\begin{table}
\centering
\caption{Continumm fluxes of centimeter sources in G035.02$+$0.35. \label{fluxes}}
\begin{tabular}{c c c c c}
\hline \hline
  Component   &    Band     &      R.A.\,(J2000)    &         Dec.\,(J2000)        &  S$^{uv}_{int}$  \\
                      &                &          (h m s)          & ($^{\circ}$\,$'$\,$''$)   &      (mJy)            \\
\hline
 & & & & \\
CM1 &  C   & 18:54:00.492 & 02:01:18.34 & 12.092 / 12.747 \\
       &  Ku  & 18:54:00.492 & 02:01:18.34 &  12.432 \\
       &  K   &  18:54:00.492 & 02:01:18.34 &  11.837 \\
CM2\tablefootmark{a} &  C   & 18:54:00.648 & 02:01:19.36 &  0.825 \\
       &  Ku  & 18:54:00.649 & 02:01:19.42 &  1.584 \\
       &  K   &  18:54:00.648 & 02:01:19.42 &  2.036 \\                
CM3 &  C   & 18:54:00.764 & 02:01:22.96 &  0.474 / 0.337 \\
       &  Ku  & 18:54:00.764 & 02:01:22.90 &  0.276 \\
       &  K   &  18:54:00.764 & 02:01:22.96 &  0.264 \\  
CM4\tablefootmark{b} &  C   & 18:54:00.528 & 02:01:15.46 &  0.061 / 0.054 \\
       &  Ku  & 18:54:00.532 & 02:01:15.52 &  0.035 \\
       &  K   &  18:54:00.536 & 02:01:15.40 &  0.032 \\
CM5 &  C   & 18:54:00.508 & 02:01:21.10 &  0.061 / 0.049 \\
       &  Ku  & 18:54:00.508 & 02:01:21.10 &  0.042 \\
       &  K   &  18:54:00.508 & 02:01:21.04 &  0.046 \\       
CM6 &  C   & 18:54:00.592 & 02:01:17.74 &  0.262 / 0.290 \\
       &  Ku  & 18:54:00.592 & 02:01:17.74 &  0.257 \\
       &  K   &  18:54:00.596 & 02:01:17.68 &  0.199 \\
CM7 &  C   & 18:54:00.252 & 02:01:10.96 &  0.031 / 0.024 \\
       &  Ku  & 18:54:00.252 & 02:01:10.90 &  0.017 \\
       &  K   &  18:54:00.252 & 02:01:10.90 &  0.026 \\  
CM8 &  C   & 18:54:01.657 & 02:01:54.04 &  0.055 / 0.068 \\
       &  Ku  & 18:54:01.657 & 02:01:53.98 &  0.049 \\
       &  K   &  ... & ... &  $<0.045$ \\                                   
       
%  & & & & \\
\hline
\end{tabular}
\tablefoot{Column\,1 lists the centimeter continuum components detected at multiple bands, as indicated in column\,2.
Columns\,3 and~4 give the peak pixel positions of the continuum sources at each band. Based on the absolute position differences 
at different frequencies, we estimate an absolute position uncertainty of approximately $0\farcs05$. Column\,5 report the radio continuum flux 
densities measured at each frequency within a common \textit{uv}-distance range, as described in \citet[][their Sect.\,3]{Sanna2018}.
The two fluxes at C band refer to the central frequencies of 5.0 (left) and 7.0\,GHz (right).
The rms and beam size are $\sim$\,10\,$\mu$Jy\,beam$^{-1}$ and  $0\farcs3$ at each band, respectively. Estimates of the 
positional uncertainties for the centimeter sources are discussed in Appendix\,\ref{Append}.
\tablefoottext{a}{Values obtained from Table\,A.1 of \citet{Sanna2018}.}
\tablefoottext{b}{Flux densities of component CM4 refer to the main peak.}}
\end{table}
}
%-----------------------------------------------------------------------------------------------------------

%\clearpage

\section{Results}\label{results}

In Fig.\,\ref{fig1}, we present a radio continuum map at C band of G035.02$+$0.35, obtained at a resolution of $0\farcs34$. For a direct
comparison with previous observations, we adopt the same labeling of radio continuum components introduced by \citet[from CM1 to CM5]{Cyganowski2011},
and extend the source numbering to the new radio components detected in this paper (CM6, CM7, and CM8). We detect eight, distinct, radio continuum
sources at multiple frequencies above a threshold of 27\,$\mu$Jy\,beam$^{-1}$ (3\,$\sigma$); six of them are distributed within a radius of
0.1\,pc from CM2. CM2 coincides in position with the brightest millimeter peak in the region, and the richest site of molecular line emission, named
core A in \citet[their Figs.\,2 and~4]{Beltran2014}. The spectral energy distribution of CM2 has a constant spectral index of 0.69 between 1 and
7\,cm, which is consistent with thermal jet emission from a young star. Its radio luminosity, of approximately 5\,mJy\,kpc$^2$ at 8\,GHz, is indicative
of an early B-type young star, according to the correlation between radio jet and bolometric luminosities (e.g., Fig.\,6 of  Paper\,I). The
radio continuum emission from CM2 was previously presented in Paper\,I \citep[see also Fig.\,3 of][]{Cyganowski2011}. In the following,
we comment on the radio continuum components detected near to CM2. 

% CM2 is also the only site associated with 6.7\,GHz CH$_3$OH masers in the region \citep{Cyganowski2011}.

In Table\,\ref{fluxes}, we list the integrated fluxes at C, Ku, and K bands for each radio continuum component. These fluxes were computed
within a common \textit{uv}-distance range of 40--800\,k$\lambda$, as described in Paper\,I, and the C-band data were split in two sub-bands of
2\,GHz each. In the side panels of Fig.\,\ref{fig1}, we analyze the spectral energy distribution of each radio component. 

CM1 is the brightest radio continuum source in the region, and it is located to the west-southwest of CM2, at a projected distance of $2\farcs55$ (or 5944\,au). 
Its spectral energy distribution is characteristic of a photoionized H\,{\sc ii} region which is hyper compact, with an angular size (best fit) of $0\farcs54$ 
corresponding to a Str\"{o}mgren radius (R$_S$) of 630\,au. This size is consistent with the (deconvolved) size obtained directly by fitting the
observed image. The four data points are consistent with a homogeneous H\,{\sc ii} region model with constant
electron density and temperature, fixed to $10^4$\,K (see Fig.\,\ref{fig1}). This model implies a number of Lyman photons (N$_{Ly}$) of 10$^{45.85}$\,s$^{-1}$,
and an electron density (n$_e$) and emission measure (EM) of 8.8\,$\times$\,10$^{4}$\,cm$^{-3}$ and 4.8\,$\times$\,10$^{7}$\,pc\,cm$^{-6}$,
respectively. The number of Lyman photons corresponds to that emitted by a ZAMS star of spectral type between B1--B0.5 and bolometric luminosity of
8--9\,$\times$\,10$^3$\,L$_{\odot}$ \citep[e.g.,][]{Thompson1984}. These values are consistent with those reported by \citet{Cyganowski2011} after
scaling their distance (3.43\,kpc) to the current value (2.33\,kpc).

The spectral index of components CM3 to CM7 was computed with the linear regression fit drawn in each panel of Fig.\,\ref{fig1}, where we also
report the spectral index values with 1\,$\sigma$ uncertainty. Notably, CM3 and CM4 have negative spectral slopes steeper than the
optically thin limit. Their spectra are consistent with nonthermal radiation within a confidence level of 3\,$\sigma$. Instead, the spectral
index of CM5, CM6, and CM7 is consistent with optically thin free-free radiation within 1\,$\sigma$.

In particular, the peak positions of radio components CM2, CM3, CM4, and CM6 are well-aligned on the plane of the sky. The four points fit a
straight line which is oriented at a position angle of 25.5$\degr$ with a small uncertainty of 0.6$\degr$ (dotted line in Fig.\,\ref{fig1}). CM3 
and CM4 are located each side of  CM2 at a similar distance of $4\farcs00$ (9320\,au) and $4\farcs29$ (9996\,au), respectively; CM6 is
located closer to CM2 at a distance of $1\farcs83$ (4252\,au). Unlike CM2, the radio continuum emission from CM3, CM4, and CM6 does not
coincide with compact dust emission \citep[e.g., Fig.\,2 of][]{Beltran2014}.
The linear distribution and the spectral index analysis support the following interpretation: \emph{CM2, CM3, CM4, and CM6 trace the same 
ionized jet, with CM2 pinpointing the origin where the central powering source is surrounded by partially opaque plasma, while CM3 and CM4 
arise from two shocks emitting synchrotron radiation located symmetrically with respect to CM2, and CM6 marks a shock emitting optically
thin free-free radiation located closer to CM2.}

%_____________________________________________________________
%                                              FIGURES  N.2
%-----------------------------------------------------------------------------------------------------------
\begin{figure}
%\sidecaption
\centering
\includegraphics [angle= 0, scale= 0.4]{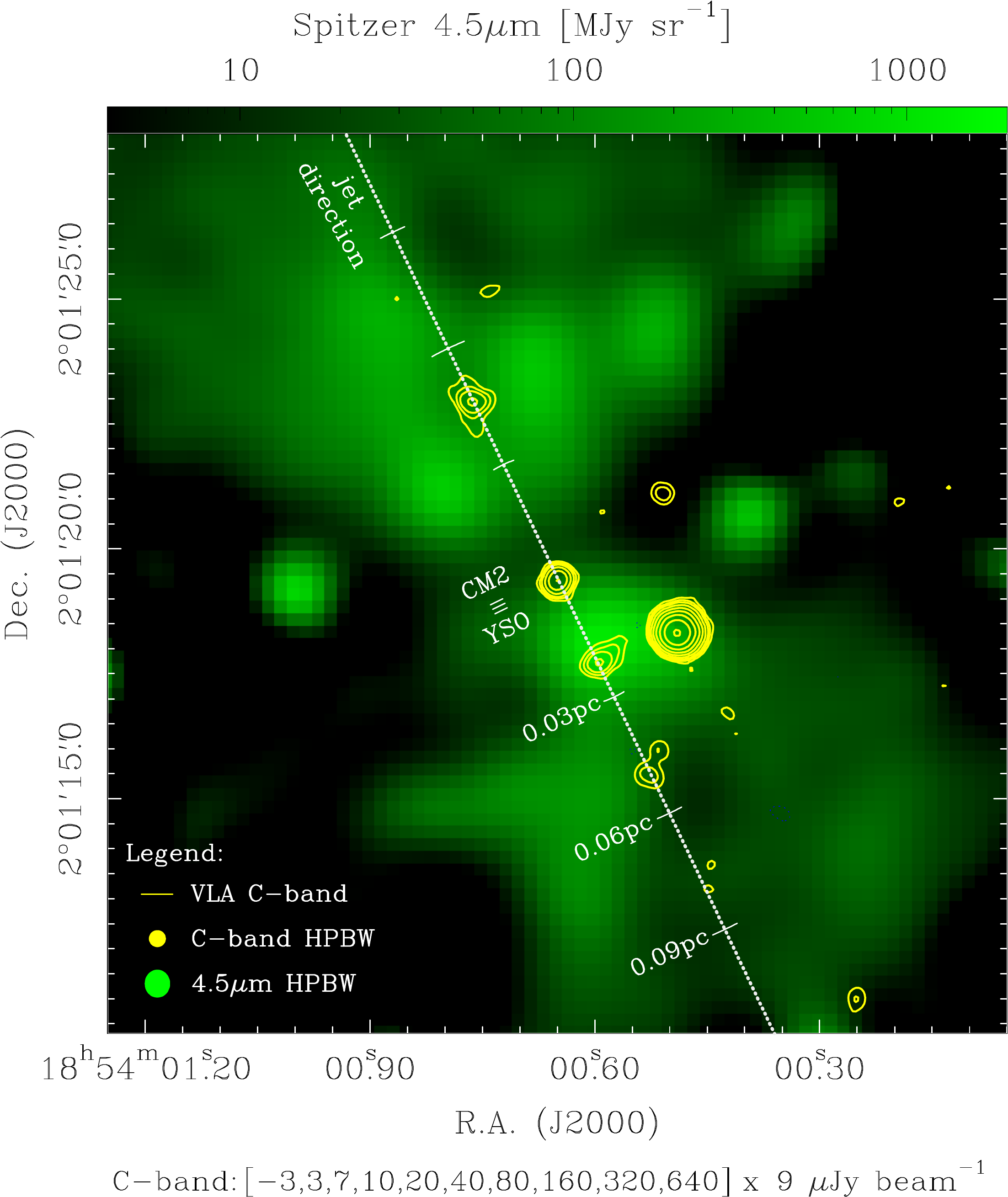}
\caption{Comparison between the radio jet emission driven by source CM2 (yellow contours) and the 4.5\,$\mu$m nebula imaged by Spitzer 
in the region (green scale). The radio continuum map is the same drawn in Fig.\,\ref{fig1}, together with the jet direction and ticks (dotted white line).
Contour levels are indicated to the bottom. The logarithmic (green) scale of the 4.5\,$\mu$m emission is drawn in the top wedge. The enhanced
resolution (HPBW) of the Spitzer map is shown in the bottom left corner (Sect.\,\ref{results} for details).}\label{fig2}
\end{figure}

%_____________________________________________________________
%-----------------------------------------------------------------------------------------------------------

%_____________________________________________________________
%                                                    TABLES # 3
%-----------------------------------------------------------------------------------------------------------
%\addtocounter{table}{0}
%\begin{landscape}
%\onltab{
\begin{table*}
%\centering
\caption{Properties of the thermal radio jet at the position of CM2.  \label{tabjet}}
\begin{tabular}{c c c c c c | c c c c c c c | r c}
\hline \hline
\multicolumn{6}{c}{Assumed parameters}  & \multicolumn{7}{c}{Radio observables} & \multicolumn{2}{c}{Jet energetics} \\  
$\epsilon$ & q$_{\rm T}$ & q$_{\rm x}$ & $F(\alpha)$ & x$_0$ & $T$ & $\nu_{m}$ & $\nu$ & $S_{8GHz}$\,$\times$\,$d^2$ & $\alpha$ & $\psi$ &
$i$ & $V_{\rm jet}$ & \multicolumn{1}{c}{$\dot{M}_{\rm jet}$}  & \multicolumn{1}{c}{$\dot{p}_{\rm jet}$}  \\
& & & & & (K) & (GHz) & (GHz) & (mJy\,$\times$\,kpc$^2$) & & ($\degr$)  & ($\degr$) & (km\,s$^{-1}$) & \multicolumn{1}{c}{(M$_{\odot}$\,yr$^{-1}$)} &
\multicolumn{1}{c}{(M$_{\odot}$\,yr$^{-1}$\,km\,s$^{-1}$)}  \\
\hline
 & & & & & & & & & & & & & &\\
1 & 0 & 0 & 1.33 & 1    & 10$^4$ & >26 & 8.0 & 1.01\,$\times$\,2.33$^2$ &  0.6   & 6 & 60--90  & 200   & 0.2$\cdot$10$^{-6}$ & 0.4$\cdot$10$^{-4}$ \\
   &   &    &          & 0.1 &              &        &       &                                           &          &    &             & 600   & 0.5$\cdot$10$^{-5}$ & 3.2$\cdot$10$^{-3}$ \\
%& & & & & & & & & & & & & &\\
\hline
\end{tabular}
\tablefoot{Columns 1 to 6 define the properties of the ionized gas following the formalism by \citet{Reynolds1986}. Those values imply a radio jet 
which approximates a conical flow ($\epsilon$\,$=$\,1), where gas has a constant (q$_{\rm T}$\,$=$\,0) temperature of 10$^4$\,K and it is
uniformly ionized (q$_{\rm x}$\,$=$\,0), with ionization fraction x$_0$. Columns 7 to 13 list the radio observables that enter into the calculation.
The last two columns report the derived jet mass loss and momentum rates.} 
%\tablefoottext{a}{} con \tablefootmark{a}
\end{table*}
%}
%-----------------------------------------------------------------------------------------------------------

In Fig.\,\ref{fig2}, we provide further evidence supporting the interpretation of a single radio jet, by comparing the spatial distribution of the radio 
continuum emission with the morphology of the 4.5\,$\mu$m emission (another typical outflow tracer). The 4.5\,$\mu$m map has been previously
presented in \citet[][their Fig.\,2]{Beltran2014}, who processed the Spitzer/IRAC data making use of the high resolution deconvolution algorithm by
\citet{Velusamy2008}. The EGO has a bipolar structure centered on CM2 and oriented in the direction of the radio continuum sources CM3, CM4, and
CM6, which, in turn, appear to bisect the EGO. Additional arguments in favor of the jet interpretation are discussed in Appendix\,\ref{Append},
based on the proper motions of CM3 and CM4 with respect to CM2, and the elongated morphology of the continuum emission. In the following, we
refer to the position angle of 25.5$\degr$ as the jet axis.

We also detect two additional radio continuum sources  farther from CM2 (in projection) along the axis of the jet, CM7 (Fig.\,\ref{fig1}) and CM8
(Fig.\,\ref{fig3}). CM7 is $10\farcs29$ (0.116\,pc) away from CM2 and only slightly offset at a position angle of 35$\degr$. CM8 is $38\farcs86$
(0.439\,pc) away from CM2 to the northeast and elongated in the direction of the jet axis. They could mark farther knots of (optically thin) ionized
gas caused by the jet passage.

The position of CM5, well away from the jet axis, argues that it traces a separate object. Its radio continuum flux, which is approximately constant between
1 and 7\,cm, is consistent with a small, optically thin, H\,{\sc ii} region (R$_S$\,$\lesssim$\,100\,au). The ionized flux can be reproduced with a
number of Lyman photons of about 10$^{43.45}$\,s$^{-1}$ \citep[e.g., Eq.\,1 of][]{Cyganowski2011}, coming from a young star with a spectral type
later than B2.5 and a bolometric luminosity of approximately 1\,$\times$\,10$^3$\,L$_{\odot}$.

\section{Discussion}\label{jet}

In the following, we quantify the mass loss ($\dot{\rm M}_{\rm jet}$) and momentum ($\dot{\rm p}_{\rm jet}$) rates of the radio jet, as well as
estimate the magnetic field strength (B$_{\rm min}$) at the position of the synchrotron knots.  

First, we apply Eq.\,(1) of \citet{Sanna2016} to estimate the mass ejected per unit time at the position of CM2, as a function of the flux density ($S_{\rm \nu}$),
frequency ($\nu$), and spectral index of the continuum emission ($\alpha$), the turn-over frequency of the radio jet spectrum ($\nu_{\rm m}$), the
semi-opening angle of the radio jet ($\psi$), its inclination with respect to the line of sight ($i$), and the expanding velocity of the ionized gas ($V_{\rm jet}$). We
assume that the radio jet can be approximated by a conical flow, where gas is isothermal at a constant temperature of 10$^4$\,K, and it is uniformly and fully
ionized (x$_0$\,$=$\,1). In Table\,\ref{tabjet}, we list the values used in the calculation. 

Values of spectral index and integrated flux density for CM2 at 8.0\,GHz are obtained from Table\,3 of Paper\,I. The spectral index is approximated to 
0.6 within the uncertainty of its measurement, and it is consistent with the assumption of conical flow \citep[e.g., Eq.\,5 of][]{Anglada2018}. A lower limit
to the turn-over frequency is set by the higher frequency of the K-band observations (26\,GHz), although the mass loss rate does not depend on the
turn-over frequency for $\alpha\,=0.6$. We have estimated the jet opening angle ($2\times\psi$) drawing the tangents to
the 3\,$\sigma$ contours of CM3 from the peak position of CM2. These tangents are approximately symmetric with respect to the jet
axis and define an aperture of 12$\degr$. The extended bipolar geometries of the radio continuum emission and IR nebula suggest that the jet axis is nearly
perpendicular to the line of sight, although a direct measurement is not available. We assume an inclination ($i$) in the range 60$\degr$--90$\degr$ which
changes the mass loss rate by less than a few percent. We further consider a lower limit to the jet velocity of 200\,km\,s$^{-1}$, based on the
proper motion analysis in Appendix\,\ref{Append} which provides an estimate of the shock velocities.

These values imply a jet mechanical force ($\dot{\rm p}_{\rm jet}$) at its origin of $0.4\times10^{-4}$\,M$_{\odot}$\,yr$^{-1}$\,km\,s$^{-1}$.
However, in the calculation we have followed a conservative approach, assuming that the jet is hundred percent ionized; this assumption implies a lower
limit of the mass loss (and momentum) rate which is inversely proportional to the ionization fraction. On the contrary, previous statistically studies of
radio jets assumed a low ionization fraction in the range 10--20\% \citep[e.g.,][]{Purser2016,Anglada2018}, which is supported by NIR observations
\citep[e.g.,][]{Fedriani2018}. Moreover, in the case of a massive young star undergoing an accretion burst (S255\,NIRS3), \citet{Cesaroni2018} showed
that the radio thermal jet emission is produced by a low ionization degree ($\sim$\,10\%). For this reason, in Table\,\ref{tabjet} we also estimate the
radio jet properties assuming that its gas is mostly neutral, and it is moving at high velocities of 600\,km\,s$^{-1}$. \citet{RodriguezKamenetzky2016}
showed that jet velocities of the order of 500--600\,km\,s$^{-1}$ might be needed to efficiently promote particles acceleration (and synchrotron
emission) in the shocks against the ambient medium. Under this conditions, we estimate an upper limit to the jet mechanical force of 
$3.2\times10^{-3}$\,M$_{\odot}$\,yr$^{-1}$\,km\,s$^{-1}$. This value is consistent with the relationship between mechanical force and radio luminosity
of jets reported in Fig.\,9 of \citet{Anglada2018}, where a low ionization fraction was assumed.

Second, we estimate the minimum-energy magnetic field strength, B$_{\rm min}$, which minimizes together the kinetic energy of the relativistic particles and 
the energy stored in the magnetic field. For a direct comparison of B$_{\rm min}$ in another radio synchrotron jet, we follow the same calculations of
\citet{Carrasco2010}. We make use of the classical minimum-energy formula, valid in Gaussian CGS units:
B$_{\rm min}$\,$=$\,$ [4.5 c_{12} (1+k) \rm L_R]^{2/7}(R^3\phi)^{-2/7}$,  where $L_R$ is the synchrotron luminosity measured in a spherical source of
radius R. The filling factor, $\phi$, accounts for a smaller region of synchrotron radiation with respect to the source size. The two coefficients, $c_{12}$ and
$k$, depend on the integrated radio spectrum and on the energy ratio among relativistic particles in the region, respectively  \citep[e.g.,][]{Govoni2004,Beck2005}. 
 
In Table\,\ref{Bmin}, we list the values used to calculate B$_{\rm min}$. We estimate the magnetic field strength for sources CM3 and CM4 separately, and 
obtain values of 1.3\,mG and 0.7\,mG, respectively. On the one hand, these values are a factor 6--3 times higher than the magnetic field strength (0.2\,mG)
estimated by \citet{Carrasco2010} in the radio jet HH\,80--81. These differences might be consistent with the uncertainty of the prior assumptions, such as
$\phi$ and $k$. On the other hand, since the synchrotron knots in HH\,80--81 are located 10 times further away from their exciting protostar than CM3 and
CM4 are from CM2, the magnetic field strength might reasonably be lower for that case \citep[e.g.,][]{Seifried2012}. Interestingly, in the synchrotron component 
of the jet driven by NGC6334I--MM1B, OH maser emission provides an independent Zeeman measurement of the magnetic field strength locally (0.5--3.7\,mG),
which is consistent with the values estimated above \citep{Brogan2016,Brogan2018,Hunter2018}. We further note that the 6.7\,GHz CH$_3$OH masers 
detected at the position of CM2 show linearly polarized emission, and this emission supports a magnetic field orientation aligned with the direction of the
jet axis \citep[][their Fig.\,5]{Surcis2015}. 

Although more data are required to fully confirm the proposed interpretation in terms of a single jet with synchrotron knots, overall these findings suggest
that G035.02$+$0.35 can provide a preferred laboratory for studying the role of magnetic fields in the acceleration and collimation of protostellar jets \citep[e.g.,][]{Kolligan2018}.

%Finally, we highlight that, as suggested by recent theoretical models \citep[e.g.,][]{Padovani2016}, those young stars which drive synchrotron jets provide
%a preferred laboratory for studying the production of cosmic rays in regions with high column densities. The local production of energetic particles can have,
%in turn, strong influence on the chemistry and the formation of protostellar discs \citep[e.g.,][]{Gudel2010}.

%_____________________________________________________________
%                                                    TABLES # 4
%-----------------------------------------------------------------------------------------------------------
%\addtocounter{table}{0}
%\begin{landscape}
%\onltab{
\begin{table}
\centering
\caption{Minimum-energy magnetic field strength in the radio jet.  \label{Bmin}}
\begin{tabular}{c c c c c c c}
\hline \hline
           & $c_{12}$       & $k$ & L$_{R}$                                 &  $R$  & $\phi$      & \multicolumn{1}{c}{$B_{\rm min}$}  \\
           &    ($10^{7}$) &        & ($10^{27}$\,erg\,s$^{-1}$)   &  ($10^{15}$\,cm) &            &                  (mG)                                \\
\hline
 & & & & & & \\
CM3 &  1.395   & 40   &  13.6  &  6.2   & 0.5 & 1.3 \\
CM4 &  1.398   & 40   &  1.2    &  5.4   & 0.5 & 0.7 \\
% & & & & & & \\
\hline
\end{tabular}
\tablefoot{Column\,1 specifies the radio continuum components used to estimate the magnetic field. Values of $c_{12}$ in column\,2 were
calculated from Eqs.\,(3), (16), and (17) of \citet{Govoni2004}; the coefficient, k, in column\,3  was obtained from Table\,1 of \citet{Beck2005}.
Column\,4 gives the synchrotron luminosity integrated between the limits of the C band. Column\,5 gives the approximate spherical radius of each
component, obtained from the geometrical average of their deconvolved Gaussian size. Following \citet{Carrasco2010}, we assume a filling factor
of 0.5 (column\,6). Column\,7 gives the minimum-energy magnetic field strength estimated from the formula in Sect.\,\ref{jet}.} 
%\tablefoottext{a}{} con \tablefootmark{a}
\end{table}
%}
%-----------------------------------------------------------------------------------------------------------

\begin{acknowledgements}

We gratefully acknowledge the thoughtful comments from an anonymous referee who helped improving the paper.
The National Radio Astronomy Observatory is a facility of the National Science Foundation operated under cooperative agreement by Associated Universities, Inc.
M.P. acknowledges funding from the European Unions Horizon 2020 research and innovation programme under the Marie Sk\l{}odowska-Curie grant agreement No 664931.
A.C.G. received funding from the European Research Council (ERC) under the European Union's Horizon 2020 research and innovation programme (grant agreement No.\ 743029)
\end{acknowledgements}

%---------------------------- REFERENCES ------------------------------

\bibliographystyle{aa}
\bibliography{asannaG35}

%_____________________________________________________________
%                                              FIGURES  N.3
%-----------------------------------------------------------------------------------------------------------
\onlfig{
\begin{figure*}
%\sidecaption
\centering
\includegraphics [angle= 0, scale= 0.8]{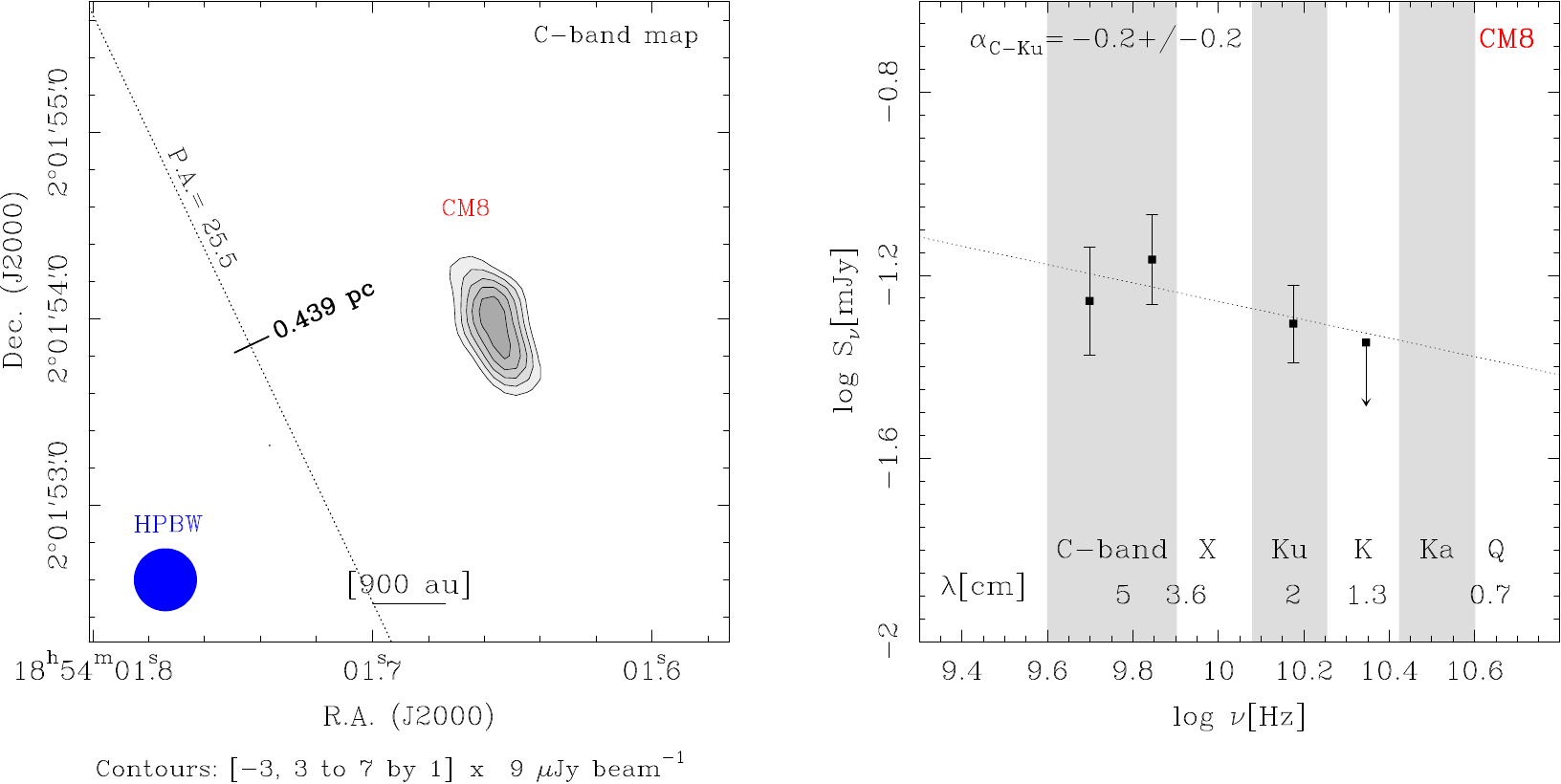}
\caption{Radio continuum map (left) and spectral index analysis (right) of component CM8, which is located to the NE of CM2, at a projected distance of $38\farcs86$.
Same symbols and labels as in Fig.\,\ref{fig1}. The spectral index of CM8 was computed with a linear regression fit between the C and Ku bands only; at K band we provide
an upper limit of 5\,$\sigma$.}\label{fig3}
\end{figure*}
}
%_____________________________________________________________
%-----------------------------------------------------------------------------------------------------------

%\clearpage

\begin{appendix}

\section{Proper motions and elongation of the radio jet}\label{Append}

%_____________________________________________________________
%                                              FIGURES  N. A1
%-----------------------------------------------------------------------------------------------------------

\begin{figure*}
%\sidecaption
\centering
\includegraphics [angle= 0, scale= 0.5]{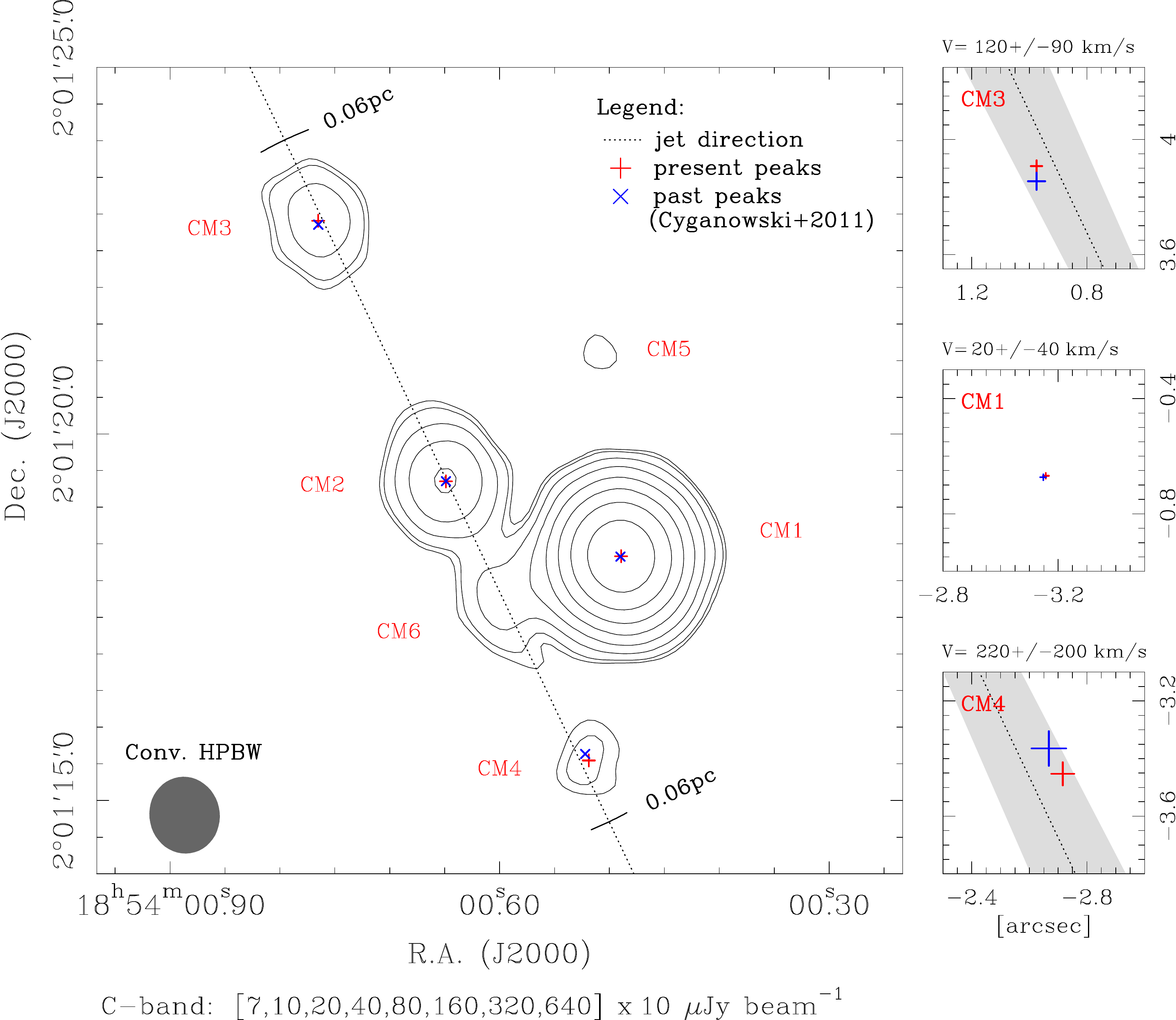}
\caption{Proper motion analysis of the radio continuum sources in  G035.02$+$0.35. \textbf{Left Panel:} 2014 C-band map (contours) processed under 
similar conditions of \textit{uv}-coverage and restoring beam size as the 2009 X-band map of \citet{Cyganowski2011}, for comparison.  Contour levels 
are indicated to the bottom at multiples of the 1\,$\sigma$ rms; the beam size is shown in the bottom left corner. Radio continuum labels, jet axis, 
and ticks are the same as in the Fig.\,\ref{fig1}. Red and blue crosses mark the centroid positions of CM1, CM2, CM3, and CM4 in May 2014 and 2009,
respectively, as obtained from a two-dimensional Gaussian fit of the C- and X-band maps \citep[X-band positions from Table\,2 of][]{Cyganowski2011}.
Individual centroids have been registered by aligning the positions of CM2 in 2014 and 2009.  \textbf{Right Panels:} zoom-in near to the centroids 
of CM3 (top), CM1(middle), and CM4 (bottom). The extent of each map is the same; angular offsets (in arcsec) are referred to the phase center.
The size of each cross quantifies the uncertainty in the centroid position (see Sect.\,\ref{Append}). For CM1, CM3, and CM4, we indicate the magnitude 
of the proper motions (and uncertainty) inferred from the displacement of their centroids between the two epochs. The dotted line and the gray shadow
mark the jet axis and its 3\,$\sigma$ uncertainty ($\pm1.8\degr$), respectively.}\label{figA1}
\end{figure*}

%_____________________________________________________________
%-----------------------------------------------------------------------------------------------------------

In this section, we compare our C-band observations with the X-band observations reported by \citet{Cyganowski2011}, with the aim
to reveal the proper motion of the radio jet and provide further proof that CM2, CM3, CM4, and CM6 belong to the same radio jet.

\citet{Cyganowski2011} observed G035.02$+$0.35 with the B configuration of the VLA at X-band (3.6\,cm) on 2009 May 7--14, 
five years before our C-band observations. In their Table\,2, they reported the centroid positions of CM1, CM2, CM3, and CM4
at that epoch, obtained from a two-dimensional Gaussian fit (CM5 was not fitted because only marginally detected). The X-band map
had a beam size of $1\farcs05 \times 0\farcs96$, oriented at position angle of $8.9\degr$, and a rms of 0.03\,mJy\,beam$^{-1}$.

To compare the C- and X-band observations directly, we have used the task \emph{clean} of CASA and imaged the C-band dataset excluding
\textit{uv}-distances greater than 300\,k$\lambda$, to approximate the \textit{uv}-coverage of the X-band dataset. We have also applied a natural
weighting and set a restoring beam size equal to that of the X-band map. The resulting C-band map (Fig.\,\ref{figA1}) well resembles that at X-band
presented in Fig.\,1 of \citet{Cyganowski2011}, where the continuum component CM6 is blended with CM1 and CM2. Notably, with respect to
Fig.\,\ref{fig1}, the continuum emission in Fig.\,\ref{figA1} shows a higher degree of elongation in the direction of the jet axis, meaning that we have
retrieved extended jet emission resolved out with the longest baselines (i.e., greater \textit{uv}-distances). In particular, the deconvolved (Gaussian)
size of sources CM2 ($\lesssim 0\farcs5\times0\farcs2$ at a position angle of $17\degr\pm10\degr$) and CM3 ($\lesssim 0\farcs6\times0\farcs1$ 
at a position angle of $21\degr\pm5\degr$) proves a strong elongation of the radio continuum emission (a factor $\geq 2.5$) and that 
this elongation is consistent with the estimated direction of the jet, within the uncertainties.

In Fig.\,\ref{figA1}, we compare the position of the continuum sources on May 2014 (red crosses) and May 2009 (blue crosses). Similar to 
\citet{Cyganowski2011}, we have fitted sources CM1, CM2, CM3, and CM4 with a two-dimensional Gaussian distribution in order to determine
their centroid positions (Table\,\ref{ppm}). We have then compared the relative position of these centroids between the two epochs
by aligning the positions of CM2, which corresponds to the origin of the radio jet.  

In the right panels of Fig.\,\ref{figA1}, we enlarge the regions around CM1, CM3, and CM4 for a detailed analysis of the centroid positions. The size 
of the red and blue crosses quantifies the positional uncertainty of the Gaussian fits; this uncertainty is calculated from the following formula:
$(FWHM/2)\times(\sigma/I)$, where $I$ and $\sigma$ are the peak intensity and the rms noise, respectively \citep[e.g.,][]{Reid1988}. The
FWHM is conservatively taken equal to the common beam size of the C- and X-band maps. These plots show that CM3 and CM4 are moving away 
from the position of CM2 and are consistent with the jet axis (dotted line) within a confidence level of about 2\,$\sigma$.
The gray shadow marks the 3\,$\sigma$ uncertainty of the jet axis. This result gains more strength when compared to the position of
CM1, which has not changed in time significantly, and further supports the intepretation that CM3 and CM4 belong to the same radio jet which
originates at the position of CM2. 

In the right panels of Fig.\,\ref{figA1}, we also report the magnitude of the proper motions for CM1, CM3 and CM4 with respect to CM2, estimated 
from the displacement of their centroids (column\,9 of Table\,\ref{ppm}). For CM3, we calculate a velocity of 120\,km\,s$^{-1}$ with an
uncertainty of 90\,km\,s$^{-1}$. For CM4, we calculate a velocity of 220\,km\,s$^{-1}$ with an uncertainty of 200\,km\,s$^{-1}$. These values are
consistent with the proper motions of nonthermal knots measured in other sources \citep[e.g.,][]{RodriguezKamenetzky2016}, providing an estimate
of the shock velocities and a lower limit on the underlying jet velocity.

%_____________________________________________________________
%                                                    TABLES # 5
%-----------------------------------------------------------------------------------------------------------
%\addtocounter{table}{0}
%\begin{landscape}
%\onltab{
\begin{table*}
\centering
\caption{Proper motion study of CM1, CM3, and CM4 with respect to CM2. \label{ppm}}
\begin{tabular}{c c l c l c c c c c c }
\hline \hline
 Component  & Epoch & \multicolumn{1}{c}{R.A.\,(J2000)} & $\Delta$x & \multicolumn{1}{c}{Dec.\,(J2000)} & $\Delta$y  & V$_{\rm x}$    &  V$_{\rm y}$  & $\rm |V|$ & P.A.  \\
    &  & \multicolumn{1}{c}{(h m s)} & ($''$) & \multicolumn{1}{c}{($^{\circ}$\,$'$\,$''$)} & ($''$) & (mas yr$^{-1}$) & (mas yr$^{-1}$) & (km s$^{-1}$) & ($^{\circ}$) \\
\hline
 & & & \\
CM1 & May 2009 & 18:54:00.49098 & 0.001 & 02.01:18.292 & 0.001 & ...  & ... & ... & ... \\
       & May 2014 & 18:54:00.48936 & 0.001 & 02.01:18.332 & 0.001 & --2\,$\pm$\,2  & 1\,$\pm$\,2 & 20\,$\pm$\,40 & $-63^{+41}_{-18}$  \\
CM2 & May 2009 & 18:54:00.6498  & 0.010  & 02.01:19.32   & 0.010 & ...  & ... & ... & ... \\  
       & May 2014 & 18:54:00.6489  & 0.006  & 02.01:19.354 & 0.006  & 0  & 0 & 0 & ... \\            
CM3 & May 2009 & 18:54:00.766    & 0.030  & 02.01:22.82   & 0.030  & ...  & ... & ... & ... \\
       & May 2014 & 18:54:00.765    & 0.010  & 02.01:22.91   & 0.010  & --1\,$\pm$\,6  & 11\,$\pm$\,6 & 120\,$\pm$\,90 & $-5^{+29}_{-27}$ \\ 
CM4 & May 2009 & 18:54:00.523    & 0.060  & 02.01:15.60   & 0.060  & ...  & ... & ... & ... \\             
       & May 2014 & 18:54:00.519    & 0.020  & 02.01:15.55   & 0.020  & --10\,$\pm$\,12  & --17\,$\pm$\,12 & 220\,$\pm$\,200 & $210^{+25}_{-43}$ \\                              
%  & & & & \\
\hline
\end{tabular}
\tablefoot{Column\,1 lists the centimeter continuum components used for the proper motion calculation. Column\,2 specifies the observational epochs.
Columns\,3 and~5 give the centroid position obtained from a two-dimensional Gaussian fit, and columns \,4 and~6 give the positional uncertainties of the 
Gaussian fits as specified in Appendix\,\ref{Append}. Positions at the first epoch are obtained from Table\,2 of \citet{Cyganowski2011}. 
Columns\,7 and~8 give the (relative) proper motion components in the east-west and north-south directions, respectively. Columns\,9 and~10 give the magnitude 
of the proper motion and its position angle (east of north).}
\end{table*}
%}
%-----------------------------------------------------------------------------------------------------------

\end{appendix}

\end{document}